\documentclass[12pt,a4paper]{article}
\usepackage{graphicx}
\usepackage{times}
\title{\bf The International \\Epsilon Aurigae Campaign 2009-2011. \\ A description of the campaign \\ and early results to May 2010
}
\author{Robin Leadbeater\\
\\
\normalsize Three Hills Observatory, UK \\ 
\normalsize robin@threehillsobservatory.co.uk
\\
\\
\normalsize Published in proceedings of \\
\normalsize"Stellar Winds in Interaction", editors T. Eversberg and J.H. Knapen. \\ 
\normalsize Full proceedings volume is available on http://www.stsci.de/pdf/arrabida.pdf
}
\date{\mbox{}}
\begin{document}
\maketitle
%
%
\def\bull{\vrule height .9ex width .8ex depth -.1ex}
\makeatletter
\def\ps@plain{\let\@mkboth\gobbletwo
\def\@oddhead{}\def\@oddfoot{\hfil\tiny\bull\quad
Workshop ``Stellar Winds in Interaction'' Convento da Arr\'abida, 2010 May 29 - June 2\quad\bull}%
\def\@evenhead{}\let\@evenfoot\@oddfoot}
\makeatother
%
%
\def\beginrefer{\section*{References}%
\begin{quotation}\mbox{}\par}
\def\refer#1\par{{\setlength{\parindent}{-\leftmargin}\indent#1\par}}
\def\endrefer{\end{quotation}}
%
%
\section{Background} 
\begin{figure}[ht]
\centering
\includegraphics[height=6cm]{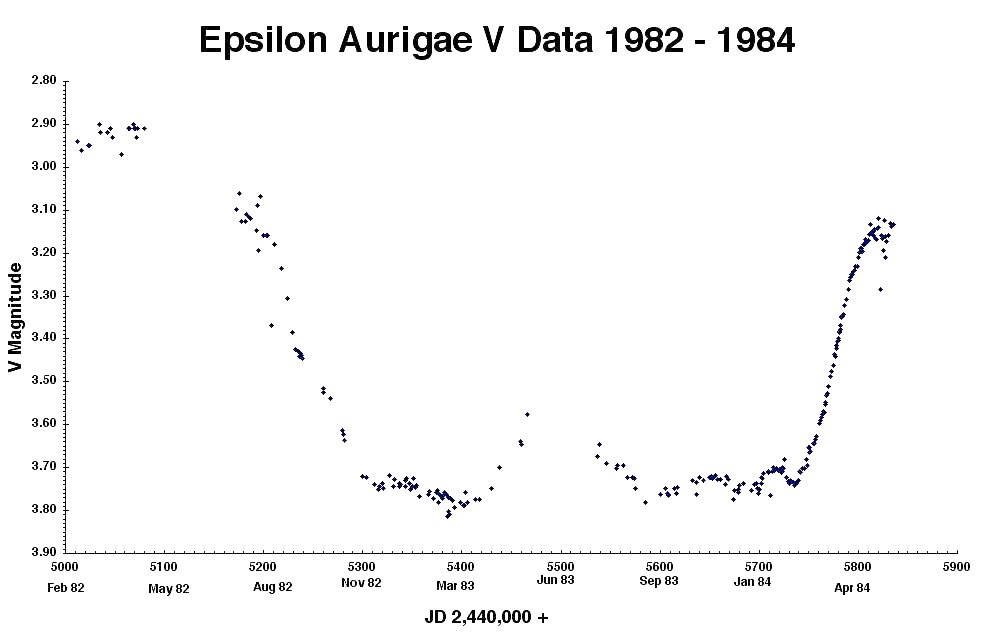}
\caption{\label{lead1}
$V$-band light curve of $\epsilon$\,Aurigae during the last eclipse (Hopkins 1987; R. Stencel, private communication).}
\end{figure}

In early 2009, immediately following the end of the WR140 periastron campaign (see these proceedings), I turned my telescope back to $\epsilon$\,Aurigae in time for the start of the eclipse. As well as being an interesting object in its own right, the Pro-Am campaign being run on $\epsilon$\,Aurigae during the current eclipse is a good example of how amateur spectroscopists can make a useful contribution.
$\epsilon$\,Aurigae is a naked eye magnitude 3 star and was first noted to be variable by Johan Frisch in 1821. It was subsequently found to be an eclipsing binary with a period of 27.1 years which undergoes an approximately 2 year long flat-bottomed eclipse  with approximately 0.8 magnitude drop in $V$ (Fig.~\ref{lead1}, note also an apparent brightening around mid eclipse in this light curve from the last eclipse.)  

All the visible light from the system appears to come from one component (an F-type star) with no visible light from the eclipsing object, which clearly cannot be a normal star given the large size and very low luminosity in the visible part of the spectrum.  Since we see only one component, there is not enough data for a complete orbital solution so there has been plenty of room for speculation over the years as to the size, mass and nature of the components and the scale of the system.  To make life more difficult, the F star also shows variability outside eclipse both in brightness and in its spectral lines.  Every 27 years a new generation of astronomers with a new generation of instruments try to solve the mystery.  This time amateurs equipped with high resolution spectrographs are joining in the fun.
%
%
\section{Review of early results from the current eclipse}

\begin{itemize}
	\item Eclipse began - 17th Aug 2009
	\item ``2nd contact"	 - 9th Jan   2009
	\item Mid-eclipse - 4th  Aug  2010
	\item ``3rd contact" - 19th Mar  2011
	\item Eclipse ends - 13th May  2011?
\end{itemize}

\begin{figure}[ht]
\centering
\includegraphics[height=6cm]{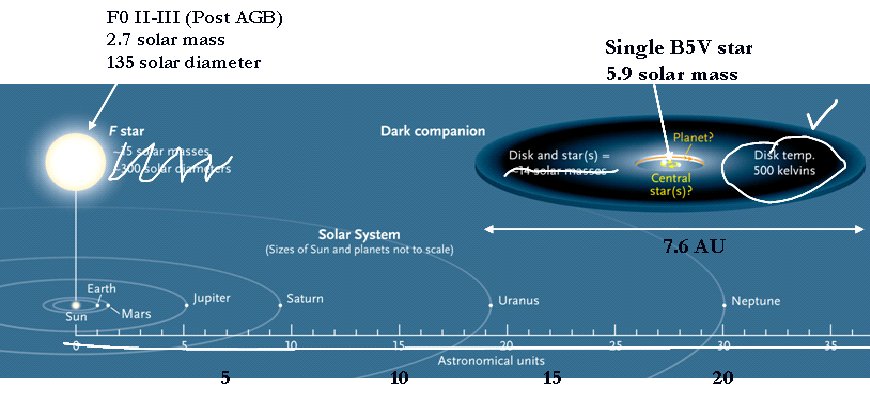}
\caption{\label{lead2}
Modifications to the consensus model following the HHS paper.}
\end{figure}

The star is currently (May 2010) in eclipse, with 1st contact  having occurred 17th August 2009 and mid eclipse due in about 2 months time (4th August 2010). 
Already during this eclipse there have been some announcements which have changed our thinking on the system. The so called ``high mass" model was the  common consensus model up to the end of last year (2009), with an F supergiant eclipsed by a disk of opaque material with possibly at least 2 B stars hidden at the centre to take care of the mass-luminosity discrepancy (Fig. \ref{lead2}).

\begin{figure}[ht]
\centering
\includegraphics[height=6cm]{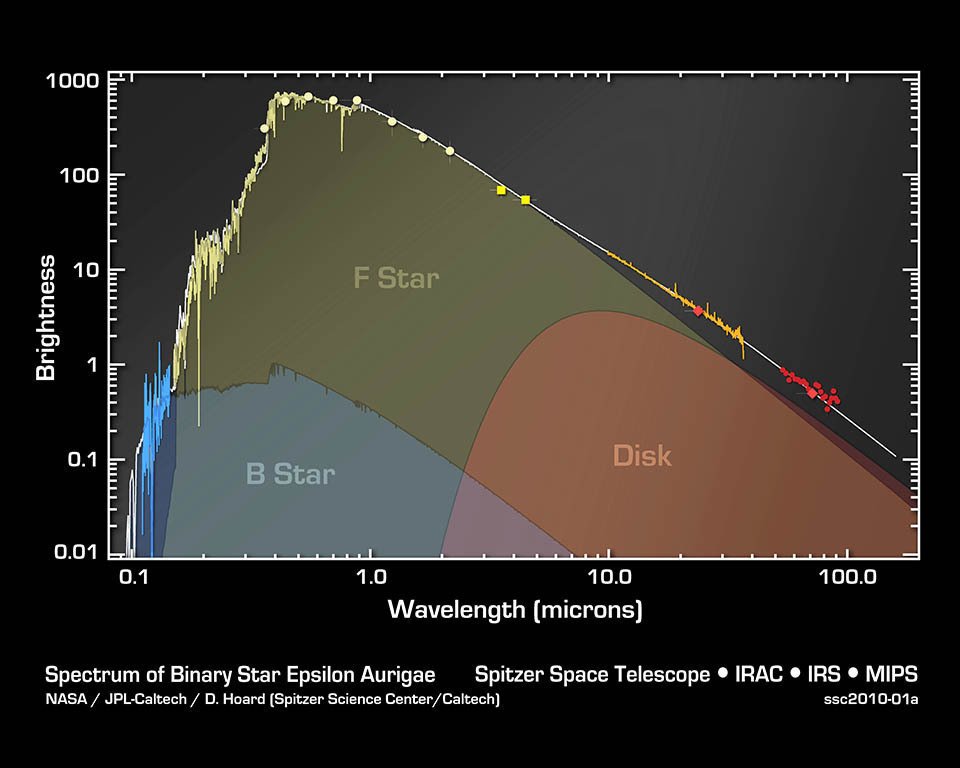}
\caption{\label{lead3}
Taming the Invisible Monster: System parameter constraints for $\epsilon$\,Aurigae from the Far-ultraviolet to the Mid-infrared (Hoard et al. 2010).}
\end{figure}

This view has now been challenged following publication of a combined UV/Visual/IR spectrum by Hoard, Howell, \& Stencel (2010; hereafter HHS) in which components from a single B5 star in the UV, a cool disk in the IR and the F star in the visible have been identified. The inference from this is that the F star is not a supergiant but a highly evolved object (post AGB?)  with low mass but large diameter. The disk would then contain a single main sequence B star. The scale of the system is also reduced by about 40\% compared with the high mass model (Figs.~\ref{lead2} and~\ref{lead3}).

\begin{figure}[ht]
\centering
\includegraphics[height=4cm]{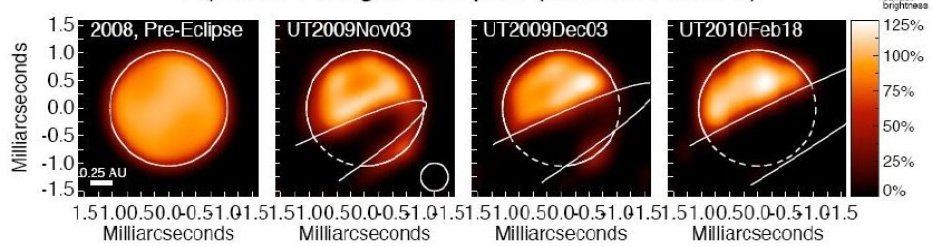}
\caption{\label{lead4}
Infrared images of the transiting disk in the $\epsilon$\,Aurigae system (Kloppenborg et al. 2010).}
\end{figure}

Also just announced in Nature this April (2010) are remarkable direct observations of the disk crossing the F star made by Kloppenborg et al. (2010) using the CHARA interferometry system (Fig.~\ref{lead4}). 
The scale of the system and rate of movement measured from the interferometry is consistent with the HHS low mass model.

\section{The role of amateurs---Photometry}

\begin{figure}[ht]
\centering
\includegraphics[height=8cm]{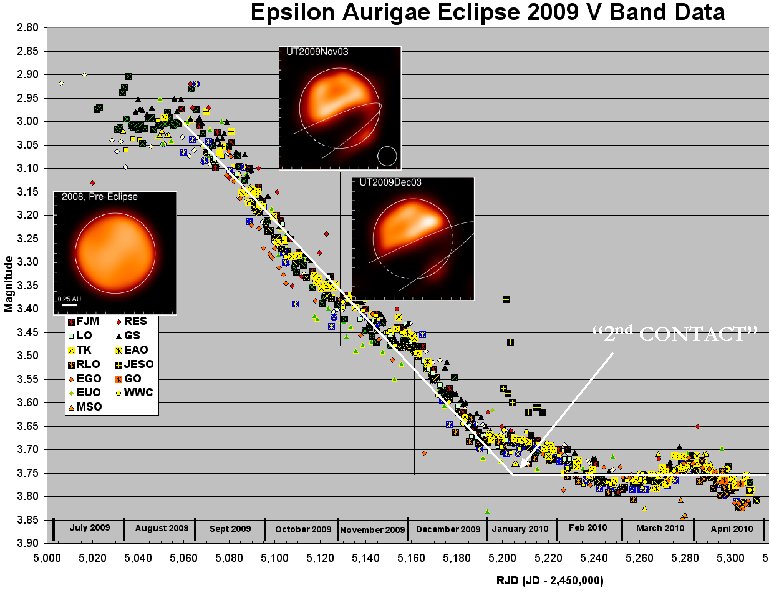}
\caption{\label{lead5}
Hopkins et al. - International epsilon Aurigae campaign 2009-11 (http://www.hposoft.com/Campaign09.html) }
\end{figure}

So what part are amateurs playing during  this eclipse? An international campaign is running to provide information exchange and to collect amateur data. This is being organised by professional Bob Stencel and amateur Jeff Hopkins, who also collaborated during the last eclipse, Jeff Hopkins coordinating the collection of photometric data. I have joined them in the past few months to coordinate the amateur spectroscopy side of things. (There is also an international outreach/education project ``Citizen Sky" being run by AAVSO, aimed at non-specialists interested in finding out how scientific research is done but I will be  concentrating on the work being done by experienced amateurs in this paper.)

By filtering out the medium-timescale brightness variations which are also seen outside eclipse, an estimate of 2nd contact can be made from the $V$ mag curve for the eclipse to date (Fig.~\ref{lead5}).  

If the CHARA images are superimposed, however, it is immediately clear that the  point identified as 2nd contact is not the traditionally defined point when the disk reaches the far edge of the star. Instead it likely corresponds to the point when the bottom half of the star is completely obscured. This explains the anomalously long ingress period ($\sim$140 days from the photometry compared with $\sim$90 days based on the star diameter and the orbital velocity of the system).

\section{The role of amateurs---Spectroscopy}

What about $\epsilon$\,Aur spectroscopically during eclipse? There are currently 12 amateur observers from five countries contributing spectra (Table~\ref{sources}). Most are following specific lines at $R\sim$18000 using the LHIRES~III spectrograph but two observers are providing wide wavelength coverage at $R\sim$12000 using  the new eShel fibre fed echelle spectrograph (see T.~Eversberg, these proceedings).  

\begin{figure}[ht]
\centering
\includegraphics[height=5cm]{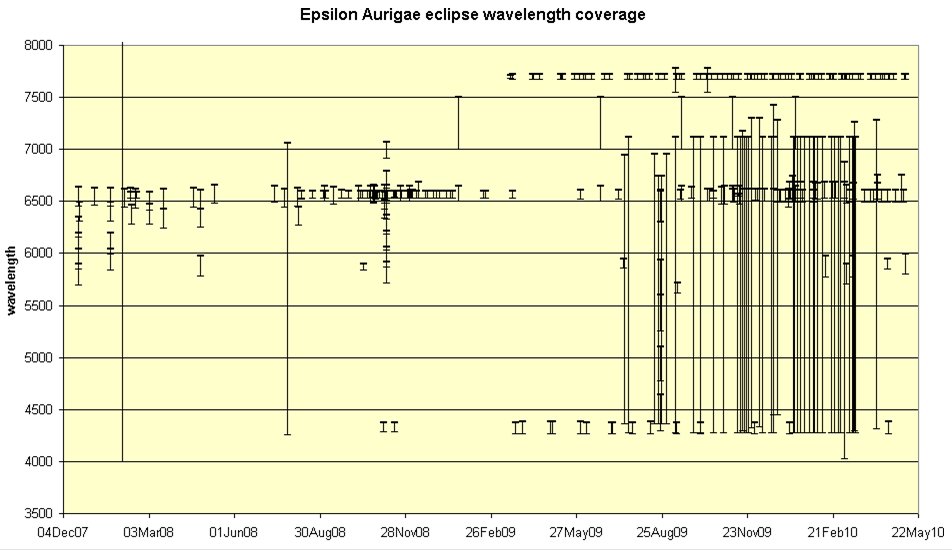}
\caption{\label{lead7} Spectroscopic coverage by amateurs during the current eclipse.}
\end{figure}

I have recently set up a web page giving access to the spectra\footnote{www.threehillsobservatory.co.uk/espsaur\_spectra.htm} 
and to date (May 2010) there are over 330. 
All the amateur data are made freely available for research purposes with conditions similar to those of the AAVSO, i.e., the campaign should be mentioned in papers and the observers included as co-authors if their data form a significant part of the paper. Figure \ref{lead7} shows the coverage to date. 

\begin{table}
\caption{List of amateurs involved in the $\epsilon$\,Aur campaign. \label{sources}}
\small
\begin{center} 
\begin{tabular}{c|c|c}
Lothar Schanne &	Germany &	LHIRES~III \\ \hline
Jose Ribeiro &	Portugal &	LHIRES~III \\ \hline
Robin Leadbeater &	UK &	LHIRES~III \\ \hline
Christian Buil &	France &	eShel \\ \hline
Olivier Thizy &	France &	eShel \\ \hline
Benji Mauclaire &	France &	LHIRES~III \\ \hline
Thiery Garrel &	France &	LHIRES~III \\ \hline
Francois Teyssier &	France &	LHIRES~III (low res) \\ \hline
Brian McCandless &	USA &	SBIG SGS \\ \hline
Jeff Hopkins &	USA &	LHIRES~III \\ \hline
Stan Gorodenski &	USA &	LHIRES~III \\ \hline
Jim Edlin &	USA &	LHIRES~III \\ 
\end{tabular}
\end{center}
\end{table}

The wide coverage of the observations made using the eShel echelle spectrograph stand out compared with the narrow range of the LHIRES. The main lines covered are H$\alpha$ and Sodium D. The set of results at 7700{\AA} are mine and I will talk about them in more detail.

\begin{figure}[ht]
\centering
\includegraphics[height=6cm]{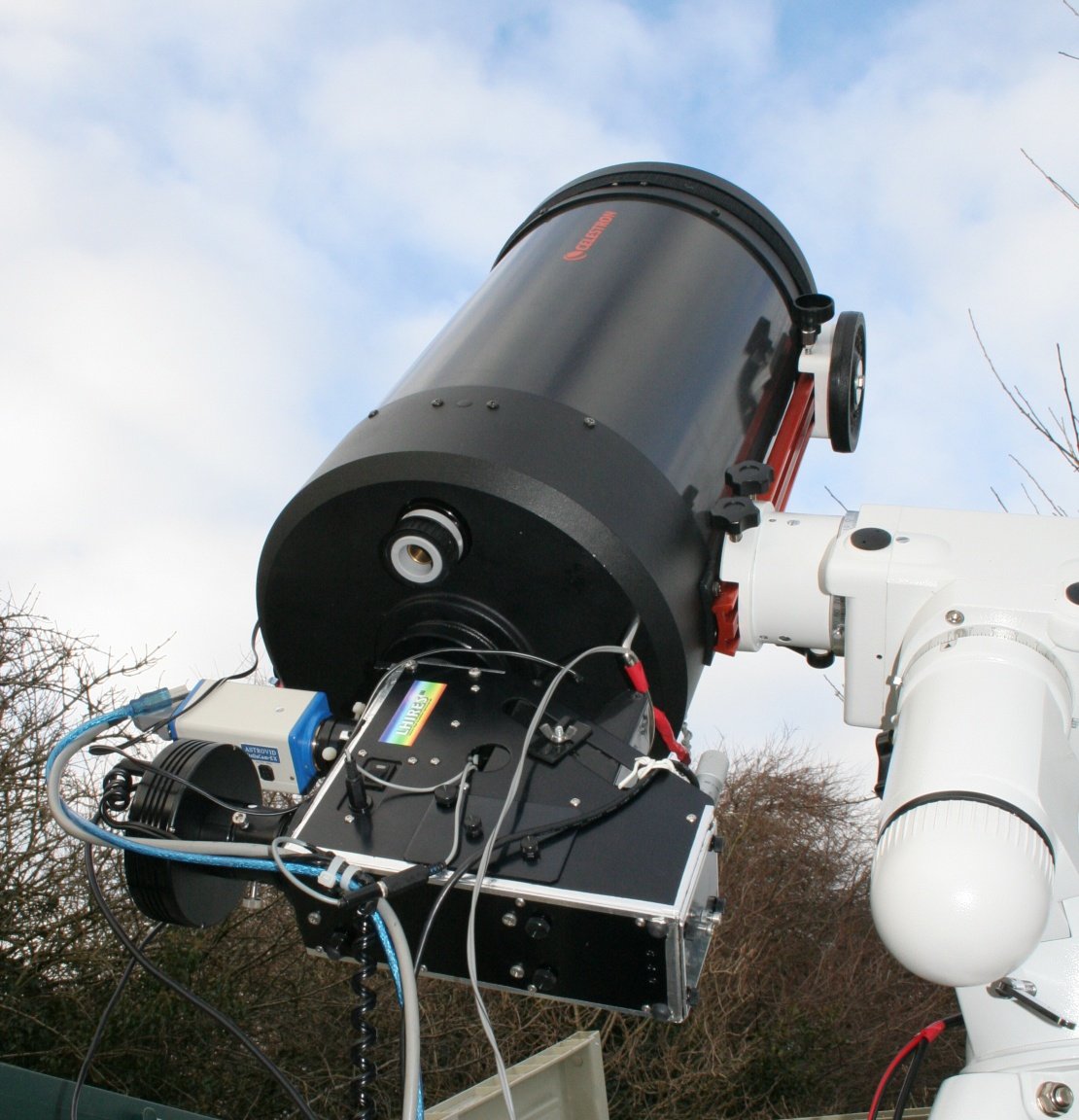}
\caption{\label{lead8}
LHIRES~III spectrograph at Three Hills Observatory.}
\end{figure}

My observatory (Three Hills - Fig.~\ref{lead8}) is located in the far north west of England and is housed in small plastic shed. The top half hinges off to reveal the telescope. It is controlled remotely from the house by wireless link. (There is no eyepiece and no
room for a human observer!)

\begin{figure}[ht]
\centering
\includegraphics[height=2cm]{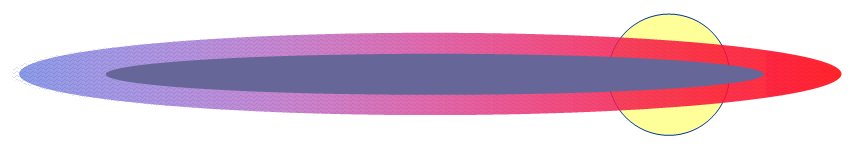}
\caption{\label{lead9}
Sketch of opaque and semi-opaque regions in the rotating eclipsing disk just after
second contact (the actual distribution of material in the disk is currently unknown).}
\end{figure}

The location is good for observing $\epsilon$\,Aur which is circumpolar from my latitude. I also have a good horizon to the north which is important as $\epsilon$\,Aur is only 8 deg altitude in June. The LHIRES~III spectrograph is attached to the 280\,mm aperture Celestron C11 telescope (this is an upgrade from the WR140 campaign when I used a 200\,mm aperture telescope.).
During eclipse the F star acts like a searchlight revealing different parts of the eclipsing object in silhouette as it moves past. This is depicted in Fig.~\ref{lead9}.  

\begin{figure}[ht]
\unitlength1cm
\begin{minipage}[t]{7cm}
\begin{picture}(7,6)
\includegraphics[width=7cm]{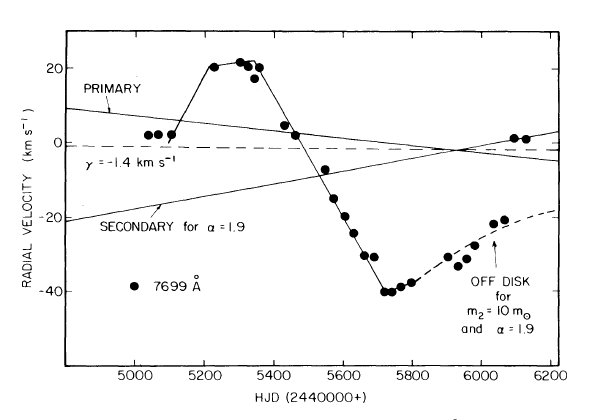}
\end{picture}
\par
\caption{\label{lead10}
K{\sc i} 7699 line radial velocity during 1982-84 eclipse (Lambert \& Sawyer 1986).}
\end{minipage}
\hfill
\begin{minipage}[t]{7cm}
\begin{picture}(7,6)
\includegraphics[width=7cm]{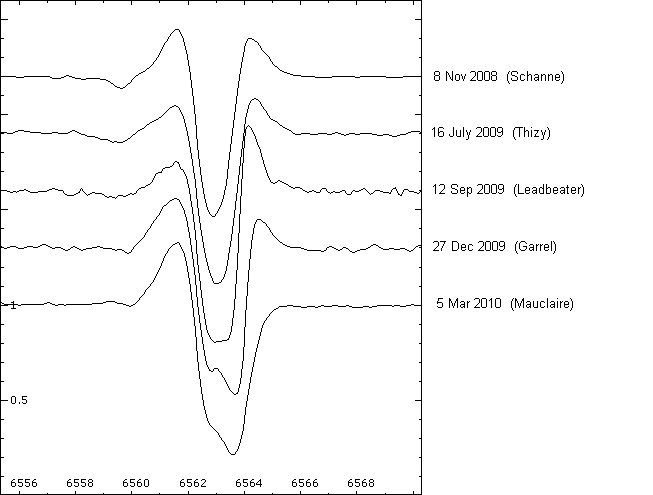}
\end{picture}
\par
\caption{\label{lead11}
H$\alpha$ line profile evolution during ingress.}
\par
\end{minipage}
\end{figure}

As well as the opaque regions which are responsible for producing  the light curve, the eclipsing object also has semi-opaque gaseous regions which produce an absorption spectrum when in front of the F star.  By measuring the intensities and radial velocities of lines in the spectrum it is potentially possible to track the distribution and velocity profile of the gas component throughout the eclipsing object as it transits in front of the F star.
The absorption features from the eclipsing disk are superimposed on the spectrum of the F star. Because the F star spectrum is variable and most of the lines are common between F star and disk, the removal of the F star features can only be approximate, but the result is a spectrum of narrow lines from the disk.
By measuring the radial velocities (RVs) of the disk lines it is possible to identify the rotation of the disk. This was done during the last eclipse, for example, by Lambert \& Sawyer (1986; hereafter L\&S) who tracked the neutral potassium line at 7699\,{\AA}. The change in RV from positive to negative from ingress to egress is clear (Fig.~\ref{lead10}).

The H$\alpha$ line is an obvious line to follow but analysis is further complicated by variable emission wings which are seen outside eclipse, assumed to come from circumstellar material around  the F star (a disk or outflowing wind?). The red and blue emission components swing up and down in intensity quasi-periodically and occasionally the absorption line has been seen to diminish significantly (Schanne 2007). Despite this, the emergence of an additional increasingly  
redshifted absorption component during this eclipse is clear (Fig.~\ref{lead11}).

\begin{figure}[ht]
\unitlength1cm
\begin{minipage}[t]{6cm}
\begin{picture}(6,6)
\includegraphics[width=6cm]{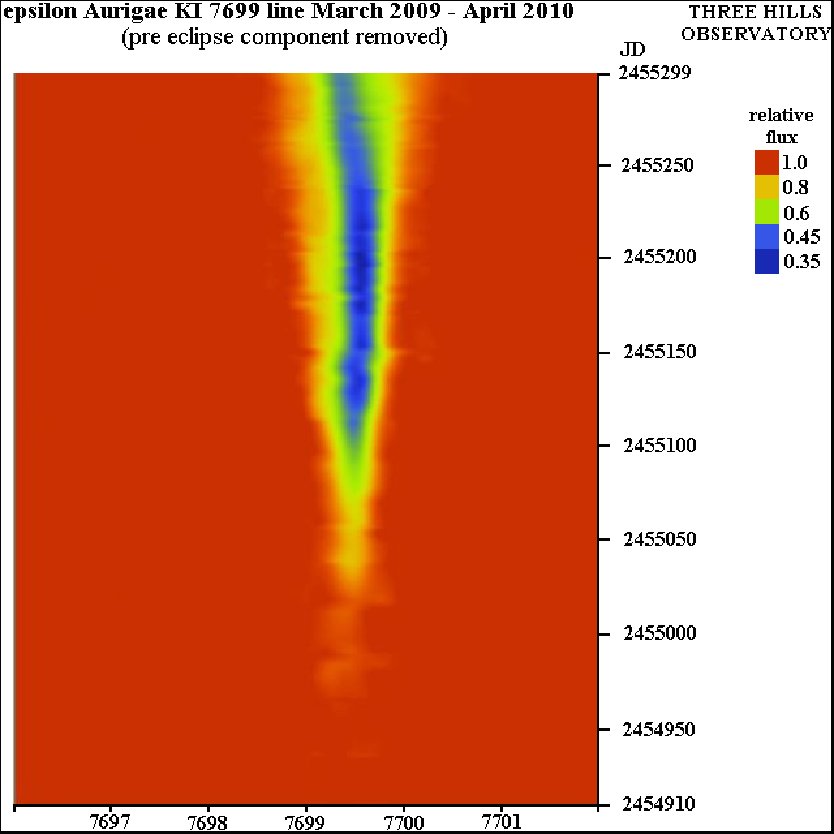}
\end{picture}
\par
\caption{\label{lead12}
7699\,{\AA} line profile evolution during ingress.}
\end{minipage}
\hfill
\begin{minipage}[t]{6cm}
\begin{picture}(6,6)
\includegraphics[width=6cm]{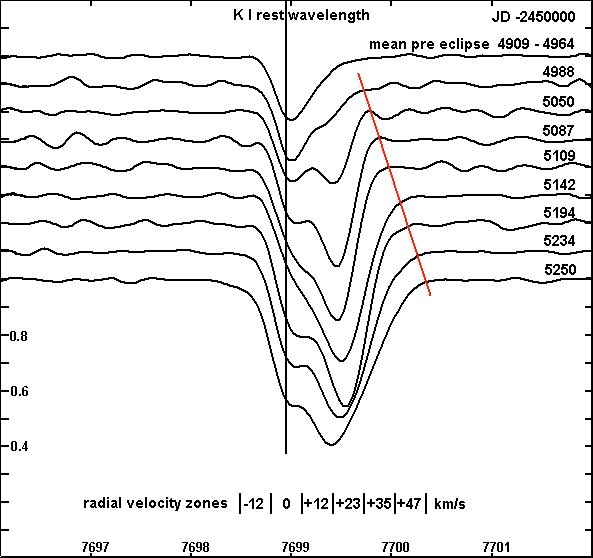}
\end{picture}
\par
\caption{\label{lead13}
Neutral Potassium 7699\,{\AA} line profile evolution during ingress.}
\par
\end{minipage}
\end{figure}

I decided to concentrate on the K{\sc i} 7699\,{\AA} line that L\&S had studied. The main reason for choosing this line is that it  is absent from the spectrum outside eclipse, so  the effect of the eclipsing object can be observed directly (there is a small constant interstellar component which can be removed). It is surrounded by O2 telluric lines, but fortunately they are far enough away not to interfere. In fact, they make useful calibration markers. To reach 7699\,{\AA} at maximum resolution I had to modify the grating adjustment on the spectrograph. At this wavelength the 2400 l/mm grating is set at a very low angle which gives good dispersion but low efficiency. 
My aim was to look in more detail at the evolution of the line profile by taking many more spectra than L\&S had been able to do (89 spectra to date, one every 4-5 days on average compared with 8 by L\&S in the same period. The false colour contour plot shows the evolution of the line (Fig.~\ref{lead12}).

Note that the eclipsing object appeared in the spectrum over 2 months before the brightness started dropping. The line gets wider and the core of the line first moves  slightly to the red and is currently (May 2010) moving to the blue.
A preliminary analysis of what has been happening in this line has been published 
(Leadbeater \& Stencel 2010). 
By measuring the maximum radial velocity at the red edge of the line we can measure the orbital velocity of the fastest rotating component in front of the F star during the eclipse (Fig.~\ref{lead13}).

\begin{figure}[ht]
\centering
\includegraphics[height=5cm]{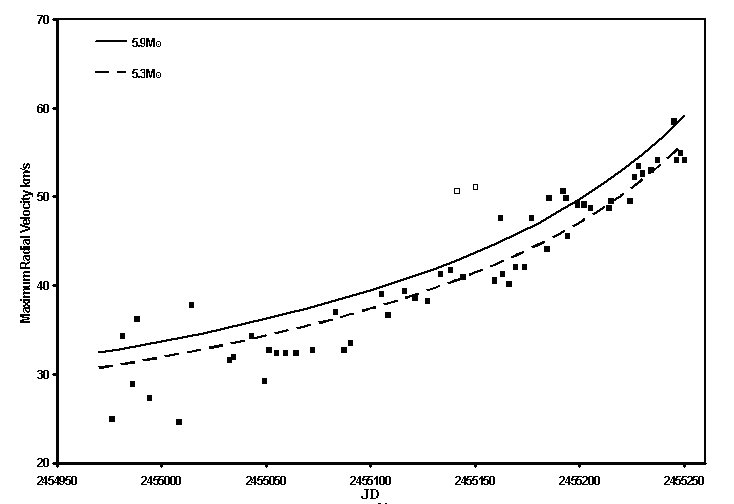}
\caption{\label{lead14}
Radial velocity of the red edge of the 7699\,{\AA} line during ingress compared with that predicted  for a Keplerian disk.}
\end{figure}

As expected for a Keplerian disk, this velocity is low at the outer edge of the disk but increases as the smaller-radius parts of the disk move in front of the F star. These data can be used  to estimate the central mass, giving a result which is consistent with the value published in the HHS paper (Fig.~\ref{lead14}).

\begin{figure}[ht]
\centering
\includegraphics[height=5cm]{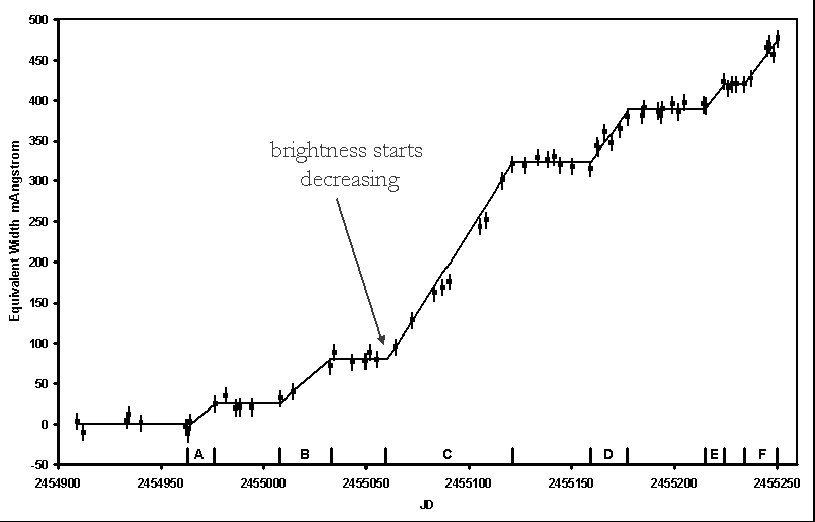}
\caption{\label{lead15}
Excess equivalent width of 7699\,{\AA} line during ingress (out of eclipse component removed).}
\end{figure}

By plotting the equivalent width of the eclipsing object component of the line, we see that the development is not smooth but takes place in a series of steps, marked by letters (Fig.~\ref{lead15}, the first two of which occur before the decrease in brightness).  

\begin{figure}[ht]
\centering
\includegraphics[height=3cm]{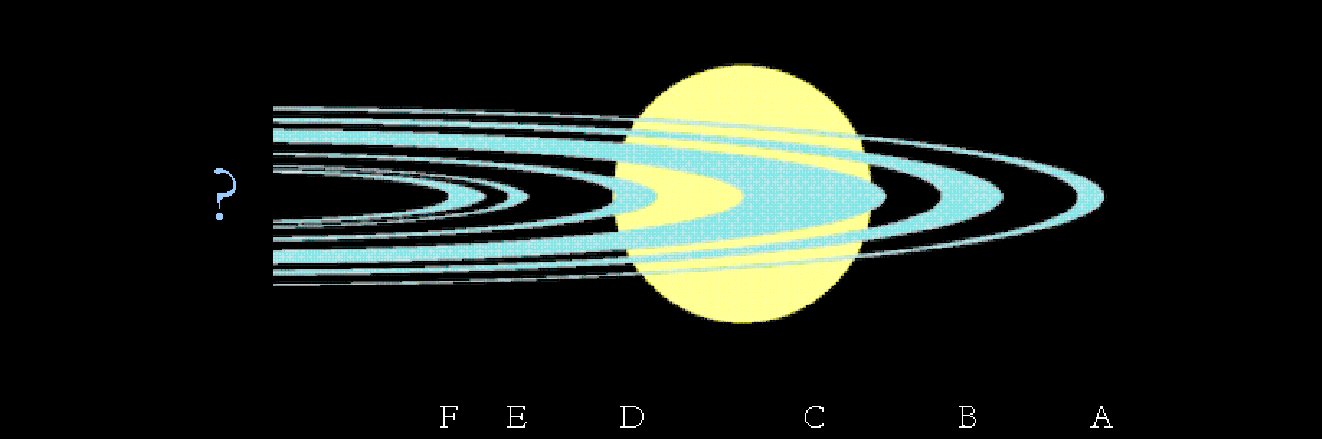}
\caption{\label{lead16}
Schematic of the postulated density variations in the eclipsing disk.}
\end{figure}

These steps were confirmed by some observations made by Apache Point Observatory
using the 3.5\,m ARC telescope (W.~Ketzeback, private communication), though they were
unable to cover the initial emergence of the K{\sc i} component. The steps are assumed to represent changes in density in the disk profile and it is difficult to resist speculating that the disk might have a ring-like structure, though these may be spiral arms or arcs (Fig.~\ref{lead16}).

\begin{figure}[ht]
\unitlength1cm
\begin{minipage}[t]{6cm}
\begin{picture}(6,6)
\includegraphics[width=6cm]{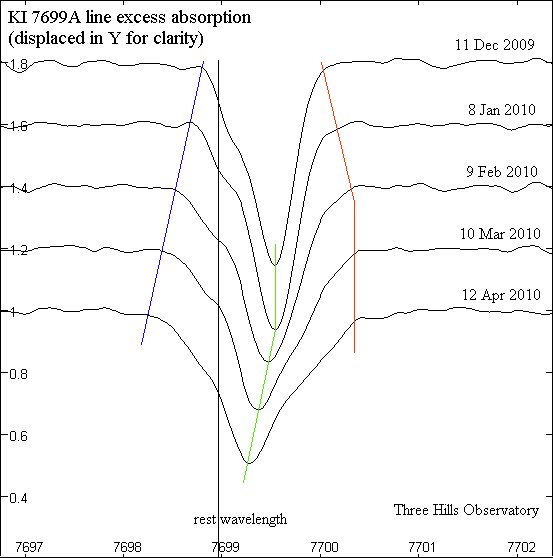}
\end{picture}
\par
\caption{\label{lead17}
Latest K{\sc i} 7699 spectra.}
\end{minipage}
\hfill
\begin{minipage}[t]{6cm}
\begin{picture}(6,6)
\includegraphics[width=6cm]{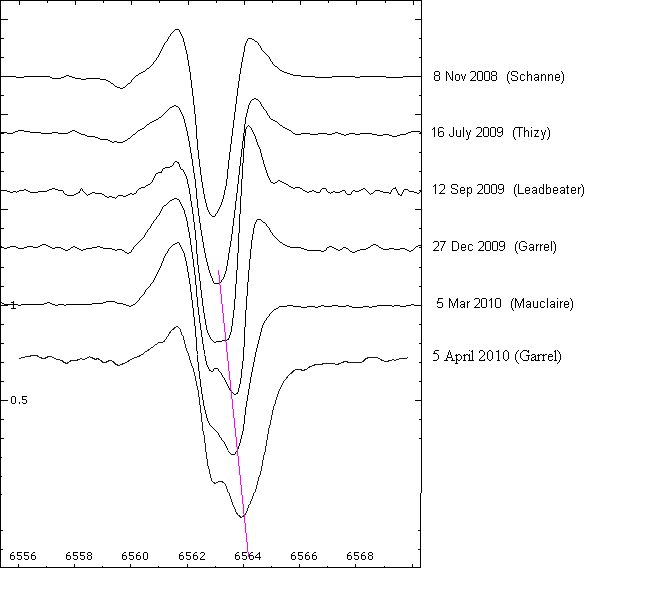}
\end{picture}
\par
\caption{\label{lead18}
Latest H$\alpha$ spectra.}
\par
\end{minipage}
\end{figure}

\section{Latest results}

\begin{figure}[ht]
\centering
\includegraphics[height=6cm]{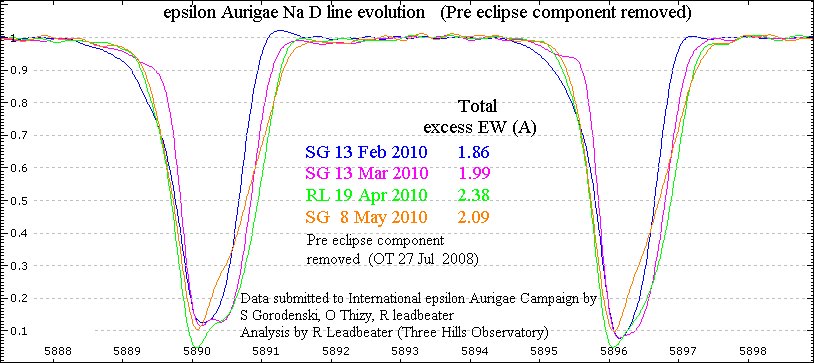}
\caption{\label{lead20}
Latest Na D spectra.}
\end{figure}

There have been differences in the behaviour of the lines on the approach to mid
eclipse. The K{\sc i} 7699 line started moving to the blue in March followed by the Na D lines
in May but as of May 2010 the H$\alpha$ line continues to move to the red (Figs.~\ref{lead17} to \ref{lead20}).

\section{Prospects for the rest of the eclipse}

\begin{figure}[ht]
\centering
\includegraphics[height=6cm]{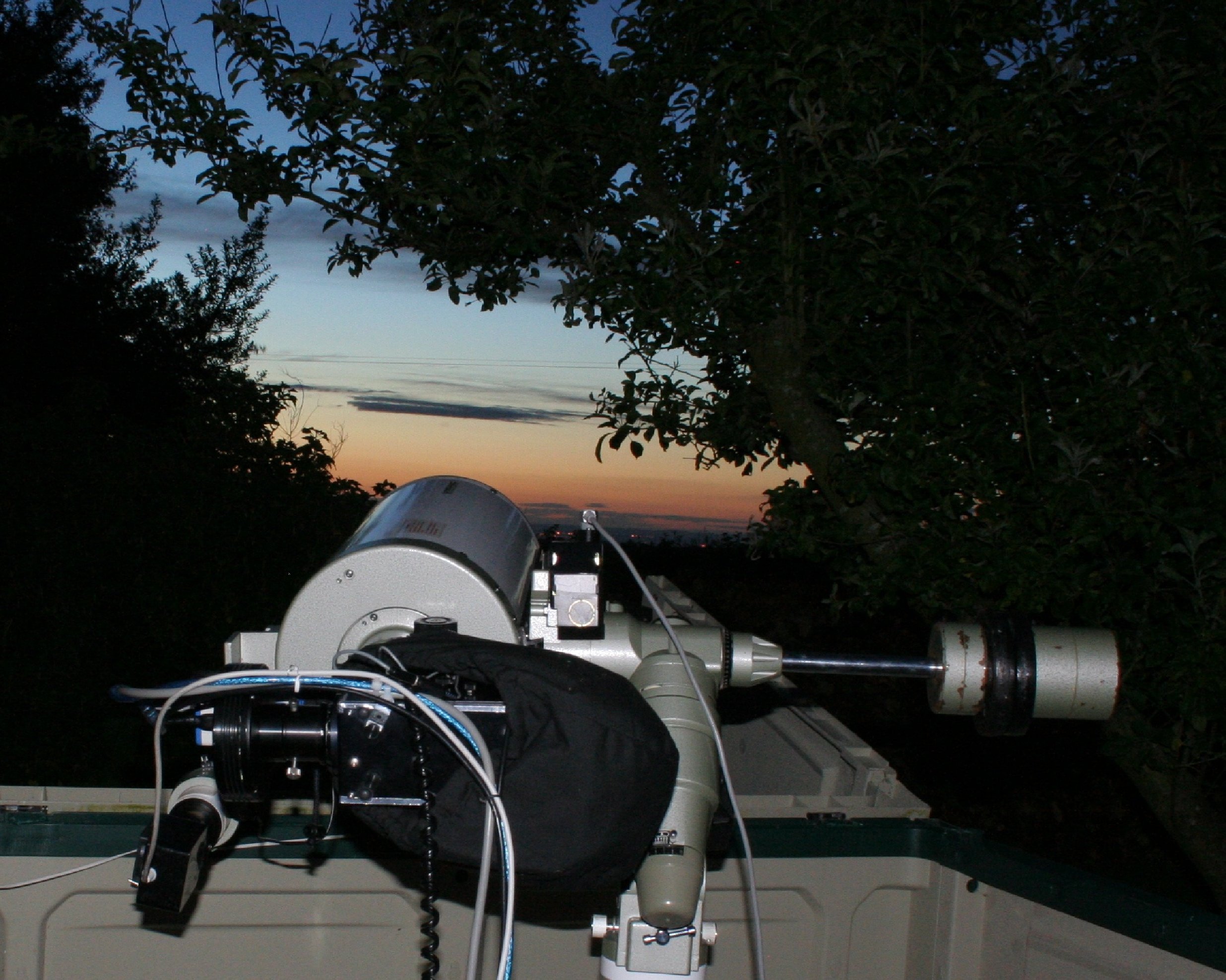}
\caption{\label{lead21}
Observing $\epsilon$\,Aurigae at Three Hills Observatory during solar conjunction 2009.}
\end{figure}

To bring the story right up to date, there are also reports that the brightness is  now increasing but caution is needed to make sure the correct extinction corrections have been applied now $\epsilon$ Aur is at low elevation (not a problem with the spectra which are normalised to the continuum.)
The next month or two will be difficult observing conditions owing to the low elevation and twilight. The picture shows an actual observing run at Three Hills at solar conjunction last June (Fig.~\ref{lead21}). There are fewer tree branches this year!

%
%
 
 \section*{Acknowledgements} 
The author would like to thank Dr Bob Stencel, University of Denver and Jeff Hopkins, the organisers of the international epsilon Aurigae campaign and all observers who have submitted data to the campaign. Thanks also to the epsilon Aurigae spectral monitoring team at Apache Point Observatory (W. Ketzeback, J.Barentine, et al.) for allowing us to view their 7699A line data.

\footnotesize
\beginrefer

\refer Hoard, D.~W., Howell, S.~B., \& Stencel, R.~E.\ 2010, ApJ, 714, 549

\refer Hopkins, J.~L.\ 1987, IAPPP, 27, 30

\refer Kloppenborg, B., et al.\ 2010, Nature, 464, 870 

\refer Lambert, D., Sawyer, S., 1986, PASP, 98, 389

\refer Leadbeater, R., \& Stencel, R., 2010, arXiv:1003.3617

\refer Schanne, L., 2007, IBVS, 5747, 1

\endrefer

     \newpage
     \qquad
          
\end{document}